# Analyzing Traffic Problem Model with Graph Theory Algorithms

Yong Tan
Independent Author
Guangzhou, Guangdong, CHINA
entermarket@163.com

*Abstract*—this paper will contribute to a practical problem, Urban Traffic. We will investigate those features, try to simplify the complexity and formulize this dynamic system. These contents mainly contain how to analyze a decision problem with combinatorial method and graph theory algorithms; how to optimize our strategy to gain a feasible solution through employing other principles of Computer Science.

*Keywords—Traffic Problem; Graph Partition Algorithm; BOTS Algorithm; Dynamic System*

## I. Introduction

It might say this paper derives from the paper *Construct Graph Logic*[1] written by author before, which was published on site arxiv.org . In that paper, it introduces the fundamental logic structures, properties, algorithms, experiments and conclusions for graph, associated to the procedures of graph traversal, partition and coloring on any instance. Hence, it is natural step for author to use these tools to research dynamic system. Author hopes to establish a general applying framework to solve such problem usually, which is generally described to how a machine makes an optimal decision at those complex circumstances. It is now reason why we choose the traffic problem, which is familiar, easy to recognize and trouble to all of us.

At our opinion, this problem is complex, dynamic and random, presently where it has been done only as how to find a shortest path between two endpoints in a city, such as *Bellman–Ford algorithm*[2] working on fixed weighted graph.

Here, we plan to look for an approach no longer on the segments with fixed weights, which status should be decided by a lot of varied factors, known or unknown, such as the time frame in a day, the amount of car sales, oil price, weather, holiday and etc. Thus, it requires an appropriate strategy, in where fewer variables can represent the superposition effects incurred by those factors. Hence, we assert the problem might be characterized to an iterative process, so that we can round this assertion likely to investigate a solution, forward in feasible, general, dynamic, flexible and real-time.

**Overview**. In the rest of this section, Section 2 introduces these essential notions of logical data structures and two major algorithms of graph. We will discuss the concept design of solution rounding those features of the traffic problem in Section 3. Section 4 will give the approach to assess the complexity of instance and with those experiments. Then in section 5, after analyzing the complexity of objects, we are going to look for optimized strategy on those fore-mentioned conclusions. Section 6 will introduce the model on industrial level and discuss some practical problems.

Finally, we will summarize above contents and pose relevant future works.

## II. Preliminaries

We are interested in the finite and connected graph, i.e., for given a node on it such that at least there is a path from it to another. We reserve the letter *n* to denote the cardinal number of nodes set on an instance and say given a graph *G* with a collection *V* of nodes with term $|V| = n$. If we say an instance is *simple*, then it means that for adjacent a pair nodes such that there is only one pair arcs toward each other among them.

Author defined a special binary relation $\tau$ among nodes and called *Traversal Relation*[1], which is a touring potential between given two nodes on an instance, implying a direction property. It can be constituted logically by an ordered pair of nodes, e.g., $u, v \in V$ and $(u, v) \in \tau$, that says there is a path, so as to a tracer can pass through *u* node to *v* node. The fact was proved to that the relation $\tau$ is the subset of the *Cartesian product* $V \times V$.

When we study the collection $\tau$, we may note that a family in $\tau$ is an equivalent class, at where each component *s* having, for given two elements $\tau_i, \tau_j \in \tau$, such that both first nodes are same $\tau_i(1) = \tau_j(1)$. We call the new guy *Unit Subgraph*, and denote this family by *S*. Furthermore the logical structure of Unit Subgraph likes a *star-tree*. It might intuitively be drawn as a central dot to send those arrows to its neighbors, showing such directivity. In term of set theory, it is a root set times to a leaves set $R(s) \times L(s)$. Well, we have a variable *m* to label the cardinal number of leaves set $|L(s)| = m$. Similarly in the same fashion, we have another family, equivalent class $\beta$ with having $\tau_i(2) = \tau_j(2)$ in the context of $\tau$ notion.

Because with the proof of families *S* and $\beta$ being the set partition of $\tau$, naturally the relationship holds in context of set theory: $\beta = S = \tau$. We thus might establish a relationship table to record those relations among these nodes on instance. We defined two modules in the approach of exploring paths that, one enumerates nodes with *S* as entry, and the another updates the relationship table with $\beta$. A graph traversal can be viewed as a process of increasing a nodes sequence, and itself may be a process of cutting graph too.

Finally, author proved such approach can expose out all potential paths on any instance with tracer beginning at any source of instance. If the input figure is a *simple graph*, the touring would be a *Hamiltonian touring*, that game's rule is visiting each node for exactly once no through any repeated path. Author called such approach *Based on Table Search*, abbr. *BOTS*.

The algorithm *BOTS* was designed to a linear program with a *While-Loop*, including those tasks: exposing new segments, backup medium results, refreshing table and issuing these final results till no path for searching. Well, we have a variable $L_T$, named *Loops Times* to represent the amount of loops. Moreover for the final results, those paths stay in varied lengths. We justly denote the quantity of paths by letter $B$, call it *Breadth* of search. Now we can summary use the ratio of $L_T/B$ as an index to assess the explored efficiency but, really it indicates what proper or bad is for the given instance to be explored by BOTS. E.g., if the instance is a single path with all nodes being sequent on it, then the value of $L_T/B$ should be $n$. If a walker starts at the root of a completed $k$ tree, then that value should be $\log_k n$. Accordingly to a path, we normally call the quantity of traversal relations *length*, denote it by $L$. The runtime complexity of search a path on a simple graph thus can be $O(mn^2)$ with this BOTS, and the overall is $O(mn^2 B)$ [1].

*Graph partition* is a heuristic approach which partition those nodes on instance to an ordered sequence components of $V$, such component we call region denoted by $\sigma$. Based upon the context of $\tau$, for each node $v \in \sigma_i (i > 1)$ such that at least there exists a node $u \in \sigma_{i-1}$, having $(u,v) \in \tau$. Namely, for node $v \in \sigma_i$, at least there is a node $u \in \sigma_{i-1}$ with an arc point to $v$. This approach does not change the collection $\tau$ and $V$, therefore it justly transform an instance to its isomorphic. An example is the following diagram.

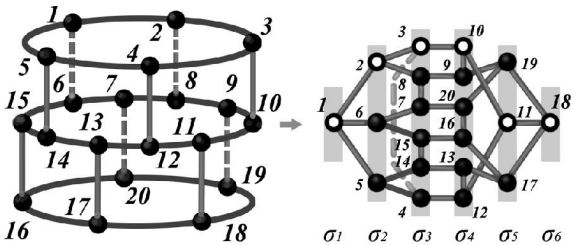

Fig. 1. *Example of graph partition*

For a simple graph, we proved the length of shortest path from source $u_{\in V}: u \in \sigma_1 (\text{for } |\sigma_1| = 1)$, to another on $i$th region $\sigma_i$ is $L = i - 1$ [1]. Here we say the shortest path between two endpoints is the minimum amount of traversal relations inserted in them. Such as node $v_1$ and $v_{18}$ on *Fig.1*, the length of path between them is $L \geq 5$, for an example path $P = v_1\ v_2\ v_3\ v_{10}\ v_{11}\ v_{18}$. In worst case, this partition algorithm has $O(n^3)$ term of runtime complexity.

Look at the node $v_3$ at the *Fig.1*, we might recognize the case of that for all leaves of it, they separately are placed in three regions: $v_2 \in \sigma_2, v_{10} \in \sigma_4$ and $v_4 \in \sigma_3$ same region is what the root $v_3$ in. That displays a filter to preprocess data of entry by using graph partitioning method, before the explored module enumerating nodes. Thus, we might encode the two methods. For all of leaves, only those in later region are worth in our attention. Of course, we might relax the restricted condition for those nodes in the root's region. Now we quickly remake an approach to specially investigate shortest paths on instance. It is clearly that two variables *Breadth* and *Length* are the central factors to affect the complexity of search. You might image these results as a *clique-tree* on a same root with $B$ and $L$ variables bounding it.

III. FEATURES AND SIMPLIFIED SOLUTION

Traffic problem is a mass to every citizen in a city. As a solution, it needs to answer a common question arisen by drivers. That is how to find a top way while in traffic jam, or whether dose such path exist. Therefore, as a driver, human extremely concerns how to quickly reach his target, more than to care about how much vehicles standstill in freeways or streets.

Hence we may give these features of this problem as follow, in intuitive idea, self-evident from our observation.

*1) We can view each road fork or intersection as a node in model city.*

*2) For a road, the Expedite Index is pass-time, not the length, especially in period of traffic jam.*

*3) The Expedite Index is dynamic, i.e., while you are on the midway, those indices of later streets might have changed than the corresponding statuses at that source.*

*4) Like a physics field, said that you should have no decision to drive a vehicle turn back to the previous node or path, unless you change your mind.*

*5) Periodicity, those similar statuses always play back in the same time frame against historical period.*

Here we assume there is a database server to record delays of traffic for all roads, that is constructed for each road by *Expedite Index* and *time frame*, from which it can be queried by someone. For periodicity property, it means that we can forecast these delays on those streets by the historical data in database server. Let $P$ be a path with $|P| \geq 1$ in above-mentioned context, we can let $P = (v_1 \tau v_2, v_2 \tau v_3, \ldots, v_{j-1} \tau v_j)$. And we denote the total time spent on path $P$ by $T(P)$, at which node $v_1$ is the source. So we can give a function to express this relationship as follow.

$$T(P)_N = \sum_{i=0}^{N} t(i), \quad \text{for } N \leq |P|.$$

The variable $t(i)$ is the delay on the $i$th segment inserted in path $P$. Set an assignment function $\omega$ for this variable $t(i)$ with those results of query from database as these entries.

$$t(i) = \omega(\tau_i, T(P)_{i-1}), \quad \text{for } \begin{cases} t(0) = T(P)_0 \text{ is initial time;} \\ t(1) = \omega(\tau_1, T(P)_0); \\ \tau_i = (v_i, v_{i+1}), \\ \quad \text{for } v_i := u; v_{i+1} := v. \end{cases}$$

This is an iterative process to cast total time with $P$ as entry by starting at $T(P)_0$. While a query is sent to server at the first node $v_1$, system can obtain the initial time-stamp $T(P)_0$ from system clock. For $i$th segment $\tau_i(1 < i < N)$, after a query with $T(P)_{i-1}$ arrives server, $\tau_i$ argument's input to server. Server might return the delay $t(i)$ involved to $\tau_i$, which is decided by the time-stamp $T(P)_{i-1}$. Then the time-stamp $T(P)_i$ is cast by $T(P)_{i-1} + \tau_i$ and, itself decides the value of $T(P)_{i+1}$ at next stage. By this way, system might ultimately gain all of offset quantities along the sequence $P$, till reach the target $v_j$.

Because the all of delays on segments reference from the historical data, we can say those results centrally present the superposition affected by those known and unknown, direct and indirect factors. For a group of total times $\bigcup_{i=1}^{k} T(P)_i$, the path reflected by minimum one is the top path naturally.

The approach seems to be naive and clumsy. But if you learn how to be an eligible taxi driver in London, you may not think so. An option of training a candidate of taxi driver in London is to keep the whole-day status of London traffic in his memory. Here we only use the server to replace human memory and, make a top decision upon it.

In term of the fourth notion above mentioned, we can regard the fact that all vehicles do not repeatedly run back on those previous roads. So as to we justly concern the *simple graph* for a touring of *Hamilton* type. For any exposed path $P$, such that there is $|P| \leq n$ with respect to graph partition being the set partition of $V$. We set any region $\sigma_i$ potential having $|\sigma_i| \leq L$. Then we can update the runtime complexity term of search a path with *BOTS* linear program above mentioned to $O(mL^2)$. Hence, the runtime complexity of overall search is $O(mL^2B)$ on a given instance.

## IV. COMPLEXITY OF SOLUTION

With this in mind of general city layout, we select the grid figure as our simulative objects in our experiments. We plan to study various type of grid for finding optimal one. We reserve to denote any grid instance by $k \times m$ with $k \geq m \geq 1$. For $k \times k$ one, that is only $k = m$. Our first type is the $k \times k$ grid, where there are $k$ nodes on row or column such that $n = k^2$. Those results of $k \times k$ Expt. are listed in following **Table 1**, with $k \in \{5,6,7,8,9,10,11\}$.

TABLE I. $k \times m$ EXPT. 1

| $n = k \times k$ | $L$ | $L_T(1)$ | $B(2)$ | (1)/(2) | R.T. |
|---|---|---|---|---|---|
| $5 \times 5$ | 8 | 251 | 70 | 3.59 | 5 ms |
| $6 \times 6$ | 10 | 923 | 252 | 3.66 | 22 ms |
| $7 \times 7$ | 12 | 3,431 | 924 | 3.71 | 57 ms |
| $8 \times 8$ | 14 | 12,869 | 3,432 | 3.75 | 655 ms |
| $9 \times 9$ | 16 | 48,619 | 12,870 | 3.78 | 18 s 160 ms |
| $10 \times 10$ | 18 | 184,755 | 48,620 | 3.80 | 3 m18 s 059 |
| $11 \times 11$ | 20 | 705,431 | 184,756 | 3.82 | 48 m 91 s |

In particular we state those items in the table as the follow, which can be onto other types for latter same experiments.

1) Column n= shows the $k \times m$ of instance.
2) Column L is the length of a shortest path.
3) Column $L_T$ is the Loops Times
4) Column B is the breadth of search.
5) Column(1)/(2) is the ratio of $L_T/B$
6) Column R.T. is the actual runtime of program*

The following Fig.2 shows that how to calculate the quantity of shortest paths (these gray shadows on B figure show those partition regions). Let nodes $v_1, v_n$ respectively be source and target.

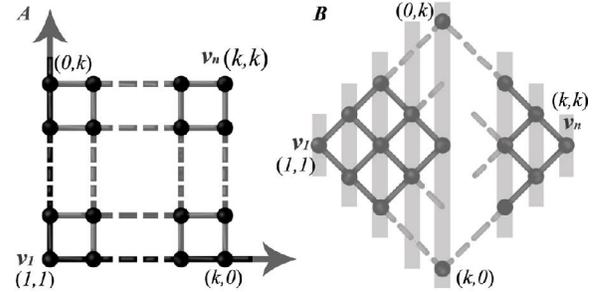

Fig. 2. *Show how to calculate a grid instance*

*B* figure shows the amount of regions is $2k − 1$ such that the length of shortest path of $v_1$ toward $v_n$ is $2k − 2$. The *A* figure shows the node $v_1$ is at the low left corner with x-y coordinates of (1, 1), and the target $v_n$ at the diagonal up right one with $(k, k)$. It is an intuitive fact that the shortest path must contain $k − 1$ segments on *x* or *y* axis respectively. Then the breadth is equal to the number of combination $S(L, L/2)$. For the length of path $L = 2k − 2$, this term may be

$$B = \binom{2k-2}{k-1} = \frac{(2k-2)(2k-3)\ldots k}{(k-1)!} \qquad (1)$$
$$= (2k-2)!/((k-1)!)^2$$

We know this is *Stirling number of the second kind* [3]. Let $s = k − 1$ for *Stirling's approximation* [4] $s! = \sqrt{2\pi s}(s/e)^s$. So that the fore term(1) with $s = k − 1$ having

$$\frac{(2k-2)!}{((k-1)!)^2} = \frac{(2s)!}{(s!)^2} = \frac{\sqrt{2\pi 2s}\left(\frac{2s}{e}\right)^{2s}}{2\pi s\left(\frac{s}{e}\right)^{2s}};$$
$$= \beta 2^{2s};$$
$$= \beta 2^{2k-2}, \qquad \text{for } \beta = \frac{1}{\sqrt{\pi s}}. \qquad (2)$$

We can set

$$\beta = 1/\sqrt{\pi k}, \quad \text{for } s \approx k \text{ and } k \gg 1.$$

We now learn the price of breadth is expensive for $k \times k$. As the term $S(n,k) = S(n, n-k)$, we can let the *m* be a constant, then having breadth $B = \binom{k+m-2}{m-1}$. We can intuitively consider that it might reduce the breadth for term(1). We firstly let $m = 4$ and $k \gg m$, these results in following *Table 2*.

---

* The tests were executed on one core of an Intel Dual-Core CPU T4400 @ 2.20GHz 2.20GHz, running Windows 7 Home Premium. The machine is 64bit system type, clocked at 800MHz and has 4.00 GB of RAM memory. The executive program is on console, which was edited and compiled by C++11, Code::Blocks 12.11 IDE; http://www.codeblocks.org

TABLE II. $k \times m$ EXPT. 2

| $n = k \times 4$ | $L$ | $L_T(1)$ | $B(2)$ | (1) | R.T. |
|---|---|---|---|---|---|
| $25 \times 4$ | 27 | 2,3750 | 2,925 | 8.12 | 1 s 399 ms |
| $30 \times 4$ | 32 | 46,375 | 4,960 | 9.35 | 3 s 591 ms |
| $35 \times 4$ | 37 | 82,250 | 7,770 | 10.59 | 9 s 565 ms |
| $40 \times 4$ | 42 | 135,750 | 11,480 | 11.82 | 28 s 034 ms |

Item $B(2)$ shows the breadth equal to $S(k+2,3)$ that is the term $S(k+m-2, m-1)$ for $m=4$. We compare two instances $25 \times 4$ and $10 \times 10$, both $n$s are equal. Observe that the length of path on $25 \times 4$ increases 50% than $10 \times 10$, but for search breadth is 6% of $10 \times 10$. And the runtime of $25 \times 4$ is below 0.5% against rival, but the efficiency of search reduces 50%. We set $m = 3$ and continue to observe the results as follow.

TABLE III. $k \times m$ EXPT. 3

| $n = k \times 3$ | $L$ | $L_T(1)$ | $B(2)$ | (1)/(2) | R.T. |
|---|---|---|---|---|---|
| $55 \times 3$ | 56 | 30,855 | 1,540 | 20.04 | 751 ms |
| $60 \times 3$ | 61 | 38,710 | 1,830 | 21.15 | 327 ms |
| $65 \times 3$ | 66 | 50,115 | 2,145 | 23.36 | 436 ms |

The breadth is $S(k+1, 2)$. We seem to consider that it is the best solution for reduce the value of $m$. We will study the time complexity to confirm this assumption.

## V. TIME COMPLEXITY OF SEARCH

It is out of the question for everyone to take a long time in cabs, only to wait for a result of top path. We aim to find an optimal type of figure to guarantee the associated computation halt in a reasonable time, through studying runtime complexity of these fore instances. We gave the term $O(n^3)$[1] represent the runtime complexity of above-mentioned graph partition with any entry, in worst case. Accordingly, we consider the runtime complexity of solution shouldn't be greater than that term.

Firstly for each potential of leaves set in unit subgraph $s$ being $|L(s)| \in \{2,3,4\}$, we let each potential equal to 4 in the worst case. Now we discuss the entry $n = k \times k$.

For $L = 2k - 2$, such that we can obtain the runtime term for exploring a path

$$O(4(2k-2)^2) = O(16k^2 - 32k + 16)$$
$$= O(16k^2), \quad \text{for } k > 1.$$

We may have the term of runtime complexity of an overall search on $k \times k$ instance.

$$O(16k^2 \cdot \beta 2^{2k-2}) = O(4\beta 2^{2k} k^2)$$
$$= O(4\beta n 2^{2\sqrt{n}}), \quad \text{for } k = \sqrt{n}; \quad (3)$$
$$= O(4n^{3/4} 2^{2\sqrt{n}}), \quad \text{for } \beta = \frac{1}{\sqrt{\pi k}}.$$

Now if $2^{2\sqrt{n}} \approx n^{9/4}$, then we gain a time complexity term $O(4n^3)$, having

$$2^{2\sqrt{n}} \approx (\sqrt{n})^{9/2}$$
$$2\sqrt{n} \log_2 2 \approx \frac{9}{2} \log_2 \sqrt{n}$$
$$4k \approx 9 \log_2 k \quad \text{for } k = \sqrt{n}.$$

Obviously if there is $k \leq 8$, then we have $4k \leq 9 \log_2 k$ such that $2^{2\sqrt{n}} \approx n^{9/4}$. Otherwise we have $k > 8$, thus there is the term $k \gg 2 \log_2 k$. Of course maybe you have pinned up the runtime changing in *Table 1*.

*Second* for $k \times m$ grid figure, with respect to $L = k + m - 2$, the price of exploring a path can be

$$O(4(k+m-2)^2) < O(4(2k)^2), \quad \text{for } k > m;$$
$$= O(E^2), \quad \text{for } k = n/m \text{ and } E = 4n/m.$$

Consider breadth term $S(k+m-2, m-1)$ for $m=4$, we can have the explicit form

$$\binom{k+4-2}{4-1} = \binom{k+2}{3} = \frac{1}{6}(k^3 + 3k^2 + 2k);$$
$$< \frac{1}{6}(6k^3) = k^3 = \left(\frac{n}{4}\right)^3, \quad \text{for } k = n/4.$$

Then the overall runtime complexity is

$$O(E^2 B) = O\left(\left(\frac{4n}{4}\right)^2 \cdot \left(\frac{n}{4}\right)^3\right);$$
$$= O\left(\frac{1}{2} N^5\right), \quad \text{for } N = n/2.$$

With the same fashion we can have the breadth for $m=3$ and $k = n/3$ is

$$\binom{k+3-2}{2} = \binom{k+1}{2} = \frac{1}{2}(k+1)k;$$
$$= \frac{1}{2 \cdot 3^2}(n^2 + 3n) < \frac{2}{9}n^2.$$

And the overall runtime complexity of $k \times 3$ is

$$O((4n/3)^2 \cdot 2n^2/9)) = O(2N^4), \quad \text{for } N = 2n/3.$$

Certainly, we can understand the runtime complexity of search for type $m = 2$ is $O(n^3)$. This is obviously that we cannot use such pure grid for our solution.

For above discussion, we may know there are two variables to affect overall complexity, which are the length of path $L$ and the inherent breadth $B$ of instance. They are both relevant to the structure of figure. For above term(3), we can obtain a solution of $O(n^3)$ if $k \leq 8$. Hence, we image that while we composite several such ones, we can obtain an optimized solution; don't worry about the $n$ increasing. The diagram we designed is following.

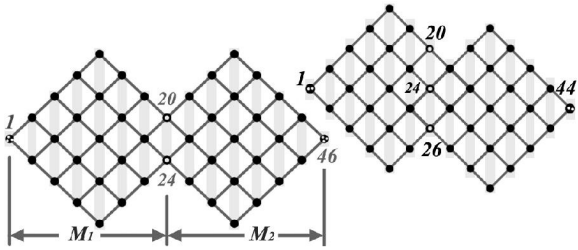

*Fig. 3. Show bridge-nodes and bridge-region*

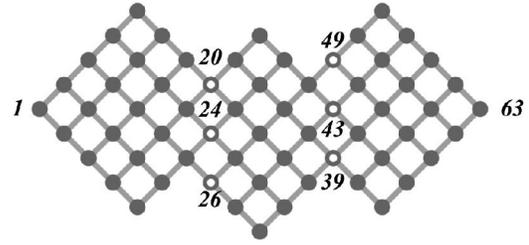

*Fig. 4. Show 3 blocks for experiment*

TABLE V.    THREE $5 \times 5$ COMPOSITING EXPT.

| $n =$ | $L$ | $L_T(1)$ | $B(2)$ | (1)/(2) | **R.T.** |
|---|---|---|---|---|---|
| 63 | 18 | 11 3051 | 31 | 3.64 | 1 s 469 ms |

They are two compositing types with two $5 \times 5$ grids. Let node $v_1$ is the source for both instances. And we set nodes $v_{46}, v_{44}$ be the targets respectively on left and right figure. You can observe there are two particular nodes on left figure, and three nodes on right one. We usually call them *bridges*.

This characteristic of bridge is given by counting these nodes in each region after graph partitioning. If there is such case $|\sigma_{i-1}| \geq |\sigma_i| \leq |\sigma_{i+1}|$, we call region $\sigma_i$ *bridge-region*. On the left figure, $\sigma_8$ is bridge-region containing $v_{20}, v_{24}$ nodes. Similarly, for $\sigma_7, \sigma_8$ at right figure, they may be bridge-region as fore condition. We select $\sigma_7$ as research object, in which there are nodes $v_{20}, v_{24}, v_{26}$. Accordingly, each node in that region is called *bridge-node*. We say that it is a *block* denoted by $M$, which is a group containing some sequent regions. In particular, for two adjacent blocks $M_j, M_{j+1}$, such that the bridge-region is their intersection, we call it *border*. An example for the left figure on *Fig. 3* above, we can have such term $\sigma_8 = M_1 \cap M_2$. It is clear that the node $v_1$ is source and the pair $v_{20}, v_{24}$ is the target in $M_1$. In turn for $M_2$, the node $v_{46}$ is target and $v_{20}, v_{24}$ are the sources. Hence the *Combinatorial Number* of left instance is

$$\binom{7}{3}\cdot\binom{7}{3}+\binom{7}{3}\cdot\binom{7}{3}=2\,450$$

Right one is

$$\binom{6}{2}\cdot\binom{6}{2}+\binom{7}{3}\cdot\binom{6}{3}+\binom{7}{3}\binom{6}{2}=1\,450$$

We can observe their breadths both are approx. to $n^2$ for $n \in \{44,46\}$. The variable $L$ is 13 and 14 respectively with $L = 2\sqrt{n}$. Thus, we seem to obtain an optimal solution for time complexity $O(16n^3)$. Those results of experiment are in following table.

TABLE IV.    TWO $5 \times 5$ COMPOSITING EXPT.

| $n =$ | $L$ | $L_T(1)$ | $B(2)$ | (1)/(2) | **R.T.** |
|---|---|---|---|---|---|
| 46 | 14 | 8 861 | 2 450 | 3.62 | 343 ms |
| 44 | 13 | 5 276 | 1 450 | 3.64 | 155 ms |

The both lengths $L$ close to type $8 \times 8$ runtime saving more than 50%. The efficiency of search $L_T/B$ does not reduce. The results encourage us, so that we continue to add the third $5 \times 5$ on right figure as following diagram.

There are 3 blocks on instance for analyzing. For nodes $v_{20}, v_{24}, v_{26}$ are in the first border, and $v_{39}, v_{43}, v_{49}$ are in second one such that we need calculate three combinatorial numbers for three stages as follow.

$$v_1 \to v_{20} \to \begin{cases} v_{39} \to v_{63} : \binom{7}{3}\binom{5}{1}\binom{6}{2} = 2625 \\ v_{43} \to v_{63} : \binom{7}{3}\binom{5}{2}\binom{6}{3} = 7000 \\ v_{49} \to v_{63} : \binom{7}{3}\binom{4}{2}\binom{6}{2} = 3150 \end{cases} \quad (4)$$

$$v_1 \to v_{24} \to \begin{cases} v_{39} \to v_{63} : \binom{7}{3}\binom{5}{2}\binom{6}{2} = 5250 \\ v_{43} \to v_{63} : \binom{7}{3}\binom{5}{2}\binom{6}{3} = 7000 \\ v_{49} \to v_{63} : \binom{7}{3}\binom{4}{1}\binom{6}{2} = 2100 \end{cases} \quad (5)$$

$$v_1 \to v_{26} \to \begin{cases} v_{39} \to v_{63} : \binom{6}{2}\binom{5}{2}\binom{6}{2} = 2250 \\ v_{43} \to v_{63} : \binom{6}{2}\binom{5}{1}\binom{6}{3} = 1500 \\ v_{49} \to v_{63} : \binom{6}{2}\binom{5}{0}\binom{6}{2} = 225 \end{cases} \quad (6)$$

We may observe that the search breadth is

$$(4) + (5) + (6) = 31\,100$$

The value is less than $2^3 \cdot n^2$ for $n = 63$. Because the length $L$ is $L = 2\sqrt{n} + 2 \approx 18$, then we compare this figure with $10 \times 10$, to which variable $L$ is approximated. The runtime on new figure is only 0.7% against $10 \times 10$.

But while we add the fourth grid, the breadth $B$ increases to 660 580 with $n = 82$. We can easily observe that the breadth greater than $82^3$. In fact, we can have the involved term $n^{N-1}$ to search breadth, which $N$ is the number of composited figures or, the amount of blocks, we call it *depth*.

Here we seem to forget our goal. Our task is to target a path onto a minimum total time. For two adjacent blocks, their border is bridge-region, more so we should employ the *Divide and Conquer* principle to finish *native-search* based on individual block. Accordingly, our optimized strategy is that: for each block, we firstly find out all shortest paths inserted in sources and targets, so that we might iteratively assign the initial time-stamps to find every top path from individual source to all target in each block, along blocks sequence. Finally, we can draw a track with minimum total delays.

For example on *Fig.4*, we may cast three top paths from $v_1$ to $v_{20}, v_{24}, v_{26}$ individually, so as to gain three time-stamps to put them on those bridge-nodes at first border. We may thus turn out 9 top paths with fore three time-stamps toward $v_{39}, v_{43}, v_{49}$ in $M_2$ with same fashion. Finally, there are three top paths between $v_{39}, v_{43}, v_{49}$ and target $v_{63}$. We can ultimately target the top path through the several stages. The breadth quantity is

1) $v_1 \to \{v_{20}, v_{24}, v_{26}\}$:

$$\binom{7}{3} + \binom{7}{3} + \binom{6}{2} = 35 + 35 + 15 = 85.$$

2) $\{v_{20}, v_{24}, v_{26}\} \to \{v_{39}, v_{43}, v_{49}\}$

1. $v_{20} \to \{v_{39}, v_{43}, v_{49}\}$

$$\binom{5}{1} + \binom{5}{2} + \binom{4}{2} = 5 + 10 + 6 = 21 \qquad (7)$$

2. $v_{24} \to \{v_{39}, v_{43}, v_{49}\}$

$$\binom{5}{2} + \binom{5}{2} + \binom{4}{1} = 10 + 10 + 4 = 24 \qquad (8)$$

3. $v_{26} \to \{v_{39}, v_{43}, v_{49}\}$

$$\binom{5}{2} + \binom{5}{1} + \binom{5}{0} = 10 + 5 + 1 = 16 \qquad (9)$$

Such that having (7) + (8) + (9) = 61;

3) $\{v_{39}, v_{43}, v_{49}\} \to v_{63}$:

$$\binom{6}{2} + \binom{6}{3} + \binom{6}{2} = 15 + 20 + 15 = 50$$

We gain the search breadth only $85 + 61 + 50 = 195$, which is approx. $3 \cdot 63$. There is $L < \sqrt{63}$ for each block. While we add a figure again, the quantity of nodes is $n = 82$ and breadth increases 61. Furthermore overall breadth should be $256 \approx 3 \cdot 82$. Because instance adds 19 nodes corresponding to adding a figure, if we have $N$ depth on instance, then the search breadth should be

$$B = 85 + (N - 2) \cdot 61 + 50 = 13 + 61N$$

And the amount of nodes is $n = 44 + 19N$. Hence, we can say that is linear growth for result $B$ to entry $n$. In the worst case, we gain a solution with runtime complexity $O(n^2)$ on the type of $5 \times 5$ compositing.

**Discussion 1.** Within the framework of above-mentioned **D&C** principle, consider an instance composited with $N$ grids. Let $n = Nk^2$. According to the term(2) of a $k \times k$ breadth, we have the search breadth for each block:

$$B' \leq \beta 2^{2k-2}.$$

In the worst case, we let length $L = 2k - a$ for each shortest path in each block, such that we have the runtime complexity term of a search for individual path.

$$\begin{aligned} O(4(2k-a)^2) &= O(16k^2 - 16ak + 4a^2); \\ &< O(16k^2), \qquad \text{for } k > a; \\ &= O(2^4 n/N). \end{aligned}$$

Hence the overall runtime complexity term is

$$\begin{aligned} T &= O(2^4 n/N \cdot NB') = O(2^4 n\beta 2^{2k-2}); \\ &= O(4\beta n \cdot 2^{2k}). \end{aligned}$$

The term is so puzzled for us to obtain such a conclusion: If $n = 2^{2k} = Nk^2$ and $k = 8$, then we gain a nice term $O(n^2)$ for such big data with having $2^{2k} = 2^{16}$ and $Nk^2 = N2^6$, such that $N = 2^{10}$. While let $N = 1$ and $k = 8$, then having $2^{2k} = 2^{16}$ that's equal to $16n^2$, we have the conclusion as previous one in native section.

This case is reasonable for practice and our experiment. It indicates that if $k$ is a constant number, then the search breadth in each block should be a constant, less than and equal to $B' = \beta 2^{2k-2}$, which is an upper boundary. If the $N \gg k$, then we have $n \gg k^2$ and $n \approx 2^{2k}$. We can write the big $O$ term as $O(B'N)$. It said that we may solve this problem with employing the **D&C** perfectly if, and only if we treat $B'$ as a constant.

**Note 1.** Consider certain root and its partial leaves, both are in a same region, likes the case in *Fig. 1*. Maybe you have paid attention to those above diagrams in native section. There is no any traversal relation in any region accordingly. It arises a problem that while we employ above approach to expose those top paths in individual block, then it is possible to double count the same segments in two blocks, for that they are at the border, intersection of two blocks.

**Summary.** This section, we attempt to find out an optimal solution round the central term of, runtime complexity. We might overcome the barrier of inherent breadth at instance, with employing the **D&C** principle. Thus, it is so bright to us: if there are $k$ machines to work for $k$ blocks respectively, they may not interfere to each other. It means that machines really implements the separate *Parallel Computing* to solve a common problem. Whenever any latter machine finishes search faster than the fore one, it absolutely might wait for event with time-stamp sent from fore one, and temporarily turn to implement other tasks.

<mark>Science and Information Conference 2015
July 28-30, 2015 | London, UK</mark>

## VI. Automated Traffic Management

At the notion of *Greedy* method, we can decide this fact: *each instance being optimal then whole being too*. Namely, if each driver obtains a top path from system, then the whole Urban Traffic should be optimized by system. So we may build the model of *Automated Management System*. We can design a *C-S* model of distributed computation, including three child-modules as follow. We show the diagram below.

*1) Client:* Terminal device, sends the query to server with source and target and receives the top path from server.

*2) Server:* Hosting historical data, recording the city map, exploring the top path for Client.

*3) Accessory Equipment:* Collecting samples of traffic in real-time so as to update database in server.

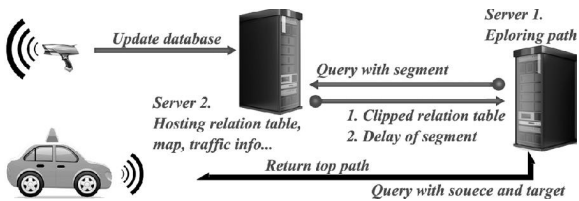

*Fig. 5. Industry model*

On level of data of city map, we need a method to clip the geographic information on it. Of course, the problem can be solved by moving a grid frame at a certain direction. Here author does not discuss this problem, because it is complex for various cities or requirements.

*Second*, it is a fact for client that the system should hold up the ability likely to deal with the traffic emergency. It says that if some accidents happen in this city and some streets in jam, system should immediately renew to feed back a new top path for each client. This time, system might regard the sudden streets as broken paths, which means them equivalent to no incident paths among involved nodes. System has to use the real-time data from *Accessory Equipment* to recast upon new relationship table. I.e., system might be still to use the rest resources to serve for client in real-time. This case is similar to one in the time frame of traffic jam. That clues system forward in the ability of flexible and real-time. If system can control the entry quantity of traffic for each street by statistics of cars or reset the limited speed on certain road, then system should carry out a dynamic management.

## VII. Summary and Future Work

Our discussions began at those abstract features of problem, and then find a solution with involved experiments for that, and then study its complexity for a best one, and then go to the second step to optimize the solution. On scholastic level, our abstracted process is $blocks \leftarrow regions \leftarrow \tau$. It made us slightly use two heuristic algorithms to finish all computing, which contains filtering useless data, exposing out the bridge-nodes and target the shortest path on instance. We ultimately find out an optimal strategy to solve this problem within polynomial time. Certainly, you maybe have more suitable circumstances as filters to optimize this problem much more.

On industry level, the traffic data can be a public resource for public transportation enterprise, such as $7 \times 24$ driverless taxi, under this automated control. Moreover, human might completely utilize these traffic data to evaluate the city layout, traffic limited speed, building road and etc.

**Future Works**. You might plan to extend above discussion to a general solution for the problem of dynamic system, which class belongs in *Decision Science* and *Operations Research*, that how to optimize a strategy or make a top decision.

Based on these abstract relationships among those nodes of factors or decisions, the process of decision may be regarded as a decision stream, which starts at premise source and, flows through those medium endpoints to the target of conclusion. In most case, it shows off an iterative process relying on an ordered sequence. Accordingly, this feature lets us can solve the class of problem with graph algorithms as fundamental framework in Computer Science.

Certainly, we might view traffic problem as a work schedule of occupying public resources. E.g., a process invokes partial CUPs for computing; robots are arranged to use elevators, channels, equipments and so on. That means such method has a wide extensive application for research in dynamic and random system.

On scholastic level, you can observe that we might have these experiments to verify some theorems of combinatorial mathematics. An example at *Fig.2*, here we briefly label a combination number for each node, which is the amount of all shortest paths from $v_1$ to that node. While you write down such combination numbers over their involved nodes, you should view the *PASCAL's triangle* appear at the left triangle on *B* figure. The nodes $\sigma_1$ and $\sigma_k$ stand on the top and bottom line of triangle range respectively. For each node in square with involved term $S(n,k)$ for $n > k > 1$, you might find two fore nodes connected to it with two terms $\binom{n-1}{k}, \binom{n-1}{k-1}$. Then we have

$$S(n,k) = S(n-1,k) + S(n-1,k-1).$$

This is a classical *Recursive formula* reflected to this experiment. The figure intuitively shows off the recursive definition and construction for *Stirling number of second kind*. Then we have a method to cast the term $S(2k,k)$ only with add operation. Hence, we have had a great approach for research of combinatorial mathematics.